# Image Prediction for Limited-angle Tomography via Deep Learning with Convolutional Neural Network

Hanming Zhang[1], Liang Li[2,3†], Kai Qiao[1], Linyuan Wang[1], Bin Yan[1†], Lei Li[1], Guoen Hu[1]

[1] National Digital Switching System Engineering and Technological Research Center, Zhengzhou, 450002, People's Republic of China

[2] Department of Engineering Physics, Tsinghua University, Beijing, 100084, People's Republic of China

[3] Key Laboratory of Particle and Radiation Imaging (Tsinghua University), Ministry of Education, Beijing, 100084, People's Republic of China

**Abstract:**

Limited angle problem is a challenging issue in x-ray computed tomography (CT) field. Iterative reconstruction methods that utilize the additional prior can suppress artifacts and improve image quality, but unfortunately require increased computation time. An interesting way is to restrain the artifacts in the images reconstructed from the practical filtered back projection (FBP) method. Frikel and Quinto have proved that the streak artifacts in FBP results could be characterized. It indicates that the artifacts created by FBP method have specific and similar characteristics in a stationary limited-angle scanning configuration. Based on this understanding, this work aims at developing a method to extract and suppress specific artifacts of FBP reconstructions for limited-angle tomography. A data-driven learning-based method is proposed based on a deep convolutional neural network. An end-to-end mapping between the FBP and artifact-free images is learned and the implicit features involving artifacts will be extracted and suppressed via nonlinear mapping. The qualitative and quantitative evaluations of experimental results indicate that the proposed method show a

---

[†] Corresponding authors: Liang Li, liliang@tsinghua.edu.cn    Bin Yan, ybspace@hotmail.com

Contacts for other authors:
Hanmung Zhang: z.hanming@hotmail.com
Kai Qiao: 15517181502@163.com
Linyuan Wang: wanglinyuanwly@163.com
Lei Li: leehotline@aliyun.com
Guoen Hu: 13838265028@126.com

stable and prospective performance on artifacts reduction and detail recovery for limited angle tomography. The presented strategy provides a simple and efficient approach for improving image quality of the reconstruction results from limited projection data.

## 1. Introduction

Limited angle tomography has gained much interest in applications like digital breast tomosynthesis [1], dental tomography [2], flat objects non-destructive inspection [3], etc. Owing to highly insufficient angular sampling, the reconstruction problem of limited angle tomography is severely ill-posed and conventional reconstruction methods, e.g. filtered back projection (FBP) and algebraic reconstruction technique [4], do not reliably facilitate satisfactory image quality or converge on the accurate solution.

In recent decades, much effect has been devoted to restraining the artifacts and improving image quality for limited angle tomography. To our knowledge, these existing works mainly focus on two strategies. One strategy is to compensate the missing sinogram via extrapolation method [5–7], another is to integrate the additional prior knowledge into reconstruction procedure. The specific prior knowledge about the unknown object, includes surfaces, density ranges or prior images, have shown helpful for artifacts suppression and edge preservation [8–10]. However, such preoperative information is often difficult to acquire and is even unavailable sometimes.

Inspired by compressive sensing theory [11], the sparse prior of images has increasingly attracted attention and the methods involving sparse regularization have been widely studied for CT image reconstruction. Total variation (TV) regularization employing the image gradient sparsity is the most popular one and shows a clear improvement in incomplete data reconstruction problems [12–18]. Besides, exemplary regularization models are higher-order derivative-based models [19–21], wavelet and curvelet-based sparse models [6, 22], and dictionary-based sparse coding models [23–25]. These sparse prior-based methods have been proven effective for incomplete reconstruction when the noise level is limited in a certain range. Unfortunately, there are various inconsistencies, such as noise, scattering, and beam hardening, in the data acquisition of actual CT systems. Reconstructing an accurate and artifacts-free image from limited-angle data is still a long-standing challenge.

To understand the limited angle problem and avoid the generation of unwanted artifacts, the characterization of limited angle artifacts has been researched. Quinto discussed the feature presentation in limited-data reconstruction and presented that the details not tangent to the projection rays will be more difficult to be recovered [26]. Based on this experience, Chen et al. proposed an anisotropic TV model to suppress the directional artifacts and obtained a better performance than the standard TV model [17]. In 2013, Frikel and Quinto derived a precise characterization of streak artifacts generated by FBP reconstruction in limited angle tomography via microlocal analysis [27]. Further, the strength of these artifacts has also been discussed and characterized mathematically [28]. These works indicated that the limited angle artifacts could be characterized specifically and may be reduced via adequate approaches.

Machine learning is a common tool to extract the signal features and generate specific learnt patterns for user demand. In medical imaging field, machine learning has played an essential role for many applications, such as computer-aided diagnosis [29–31], medical image segmentation [32], etc. An important element of machine learning is to construct features from input data. Recently, deep learning technique, which makes major advances in discovering intricate structures in high-dimensional data and improving the learning process, has been studied widely for computer vision tasks [33]. Convolutional neural network (ConvNet) and deep belief network are two primal architectures in the deep learning field as they have been well established and shown great promise for complex data processing [34].

Inspired by Frikel et al.'s work and recent development of deep learning, this paper proposed a data-driven learning method to extract and reduce the specific artifacts in the FBP reconstructions from limited-angle projections. To identify the image features efficiently, the ConvNet method with powerful capability on feature representation and intelligent learning is considered. A multi-layer ConvNet is designed to represent the features of specific artifacts in a stationary CT scanning configuration, and learn an end-to-end, pixel-to-pixel mapping between FBP reconstructions and original artifact-free images. The features involving artifacts will be extracted by feature representation and reduced by nonlinear mapping, and then a corrected artifact-less image will be predicted in the output of the network. Finally, the feasibility of our proposed method is validated qualitatively and quantitatively.

The main contribution of this work is two-fold: 1) extend the approach to study limited angle problems; 2) demonstrate that deep learning is useful in the problems of limited angle artifacts reduction, and could achieve a good quality.

## 2. Methods and experiments

### 2.1 Deep Convolutional Neural Networks

Artificial neural network doesn't need to establish accurate mathematics model, it sums up the implicit relation between systematic input and output through studying the training sample data. The convolutional neural network is a typical type of feed-forward artificial neural network and is proposed first for recognition decades ago [35]. Due to recent developments on deep learning techniques such as the nonlinear function of rectified linear unit (ReLU) [36], and the new regularization technique of "dropout" [37], ConvNets have been applied with great success in the field of computer vision.

Deep ConvNets are typically organized in a series of layers. The image is imported in the input layer, which is connected to one or more hidden layers. Linear and nonlinear filters are applied at each hidden layer to extract salient features and propagate deep information. In ConvNets, the convolutional layer and pooling layer are commonly used, and the result of weighted sum operator in convolutional layer is often then passed through a smoother non-linearity such as a tanh operator or a ReLU. Given the powerful learning capability, deep ConvNets are now the dominant approach for almost all recognition and detection tasks [33].

### 2.2 Limited-angle artifacts

In CT scanners, the object to be reconstructed is placed in the center of rotation and the x-rays pass through all parts of the object uniformly. When the system rotates regularly along a full angular coverage, the data acquired in this case is called complete tomographic data. However, there are some cases that the CT scanner cannot rotate completely around the object, and the data acquisition covers less than a 180 ° angular range. In such cases, the acquired data is called limited-angle tomographic data. As the projection data are highly incomplete, standard tomographic reconstruction algorithms, such as the well-known FBP algorithm, will not afford a reliable solution and the reconstructed image usually suffers streak and aliasing artifacts.

Consider the FBP reconstruction from limited-angle tomographic data, some phenomena can be observed in the reconstructed images: only some specific features of the original object can be recovered reliably, and some additional artifacts are generated and superimpose the reliable information [27]. Interestingly, for a stable directional angular coverage in scanning, the additional streak artifacts show a related directional property.

To understand the added artifacts mathematically, Frikel and Quinto used microlocal analysis method to characterize the limited angle artifacts and proved that the additional streak artifacts are created along lines that are tangent to singularities of the original object [27]. As shown in Figure 1, for two different limited-angle scanning configurations, though the artifacts presented in FBP reconstructions show different directional properties, the streaks correspond to the ends of the limited angular range.

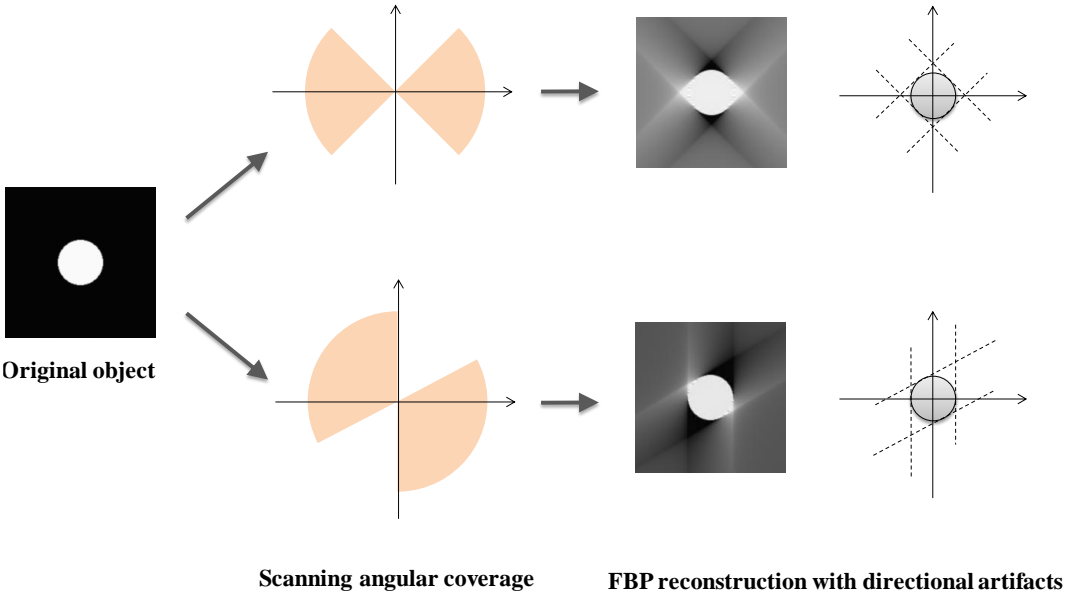

**Original object**  **Scanning angular coverage**  **FBP reconstruction with directional artifacts**

Fig. 1. FBP reconstructions on a uniform-density disk with two different cases of scanning angular ranges.

Based on the above analysis, it can be seen that the artifacts appeared in the FBP results have specific and similar characteristics in a stationary limited-angle scanning configuration.

Thus, some potential useful information under these artifacts might be exploited to improve the image quality, and it is possible to restrain the added artifacts while keep the original visible features reliably. However, the objects in practical CT scanning often have complex geometries and multiple attenuation coefficients, a precise distribution and description of the added artifacts in an FBP reconstructed image are difficult to be represented via mathematical derivation method. Therefore, our aim in this paper is to derive an artifact suppression strategy for the limited-angle tomography via a data-driven learning method.

**2.3 Deep Learning for artifacts reduction in limited-angle tomography**

For extracting and reducing the specific artifacts in FBP results, the popular deep ConvNet-based method is adopted in our design. The task of this neural network is to exploit an end-to-end mapping between low-resolution FBP images and original high-resolution images reconstructed from full-angle projections, and restrain the unwanted additional artifacts by nonlinear mapping process of learned feature representations. Through training the network with large sample data, a nonlinear prediction system will be formed to generate an image with fewer artifacts.

An overview of the proposed network is depicted in Fig. 2. In the workflow of this network, it consists of three steps:

1) Feature extraction: the first step is a feature extractor that transforms the input image into a set of feature maps;

2) Nonlinear mapping: this step maps one set of feature maps to another set of feature maps, and works for suppressing the unwanted features caused by limited-angle problem ;

3) Feature combination: the last step is a shape generator that produce predicted image via combining the features represented from nonlinear mapping procedure.

The input is an image of FBP reconstruction with pixel size $N_1 \times N_1$, and the output is the ground truth image with size $N_3 \times N_3$. For pixelwise prediction, the output size could be equal or smaller than the input size. In our implementation, $N_3$ is smaller than $N_1$ and the predicted output corresponds to the center region of the input image. There are two layers that represent the features in the network, each layer of data is a three-dimensional vector of size $C_i \times N_2 \times N_2$, where $N_2 \times N_2$ is spatial dimension, and $C_i$ is the channel dimension of the $i$-th layer.

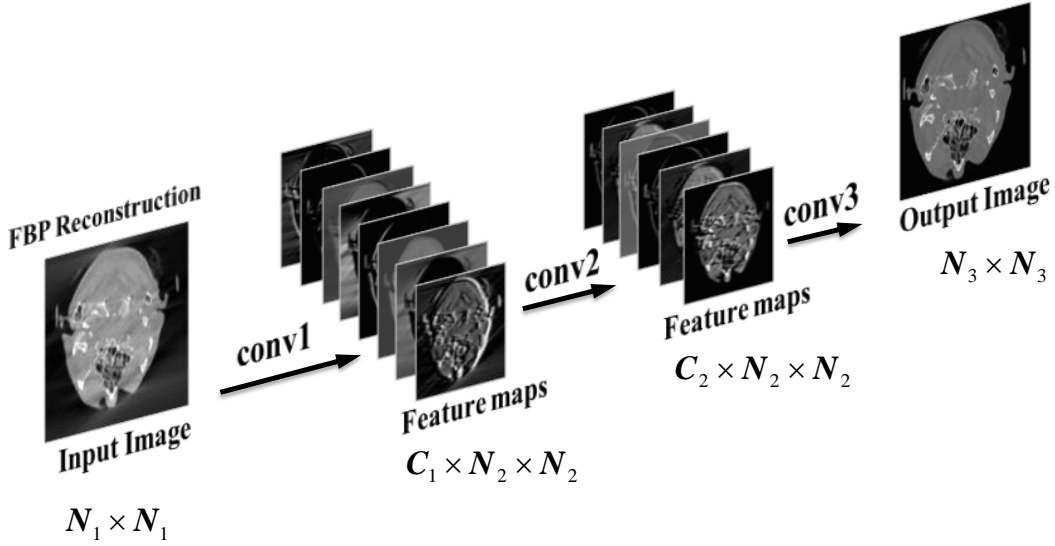

Fig. 2. Overall architecture of the proposed network.

Let $\mathbf{x}$ denote the image obtained from FBP reconstruction, and $\mathbf{y}$ denote the original image of object, the goal is to recover $\mathbf{y}$ from an blurry image $\mathbf{x}$. Thus, the input and the corresponding label (output) of the entire network are $\mathbf{x}$ and $\mathbf{y}$, respectively. Then, let $\mathbf{x}_1$ and $\mathbf{x}_2$ denote the output results of step 1 and step 2, respectively, the process of our network can be formulated as follow:

$$\text{Step 1:} \quad \mathbf{x}_1 = f_1(\mathbf{x}), \tag{1}$$

$$\text{Step 2:} \quad \mathbf{x}_2 = f_2(\mathbf{x}_1), \tag{2}$$

$$\text{Step 3:} \quad \mathbf{y} = f_3(\mathbf{x}_2), \tag{3}$$

where $f_i$ represents the system function of the $i$-th step.

Next we discuss these three steps in our network in details.

### 2.3.1 Feature extraction

Medical images contain huge amount of information and complex structural features. To analysis and distinguish the true structures and false artifacts, it needs to extract the input image to a series of elementary features, such as edges, corners, contours and so on.

In this step, convolutional operators are used to extract features. We apply $C_1$ convolutions with a kernel size of $n_1 \times n_1$ on the input image, and use ReLU non-linearity on the output of convolutional operator. Then we can get the the first layer of feature maps. The operation of step 1 could be expressed as in the following scheme:

$$f_1(\mathbf{x}) = \max(0, \mathbf{w}_1 * \mathbf{x} + \mathbf{b}_1), \tag{4}$$

where $\mathbf{w}_1$ and $\mathbf{b}_1$ stand for the weight matrix and bias of convolutional filter, respectively. $\mathbf{w}_1$ is of a size of $1 \times n_1 \times n_1 \times C_1$, and $\mathbf{b}_1$ is a $C_1$-dimensional vector.

### 2.3.2 Nonlinear mapping

Nonlinear mapping is designed for restraining the deformed features involving unwanted artifacts and correcting the aliased features to a clear one, thus the feature maps in the first layer need to be transformed to better ones. To achieve this aim, a convolutional kernel with a size of $1 \times 1$ and a ReLU non-linearity are employed here to interact and integrate information among different maps. Similar to step 1, it involves convolving the feature maps by a set of filters, and the operation could be expressed as follows:

$$f_2(\mathbf{x}_1) = \max(0, \mathbf{w}_2 * \mathbf{x}_1 + \mathbf{b}_2), \tag{5}$$

where $\mathbf{w}_2$ and $\mathbf{b}_2$ represent the weight matrix and bias, respectively. $\mathbf{w}_2$ is of a size of $C_1 \times 1 \times 1 \times C_2$, and $\mathbf{b}_2$ is a $C_2$-dimensional vector.

To explain this step clearly, we show an example in Fig. 3. An image which consists of three simple shapes is used as the scanned object. The available angular coverage of a parallel-beam CT scanning is from 0 to 150 degrees. The input of the network is an initial reconstruction by FBP method. By taking the step 1 of feature extraction, the input image is represented as a series of feature maps. Some of these feature maps have clear true characteristics, but some have aliasing and distortions. Then, these features maps are relocated and reformed through the nonlinear mapping step. It can be seen that the new feature maps show an obvious suppression on the undesired artifacts and keep the reliable characteristics at the same time.

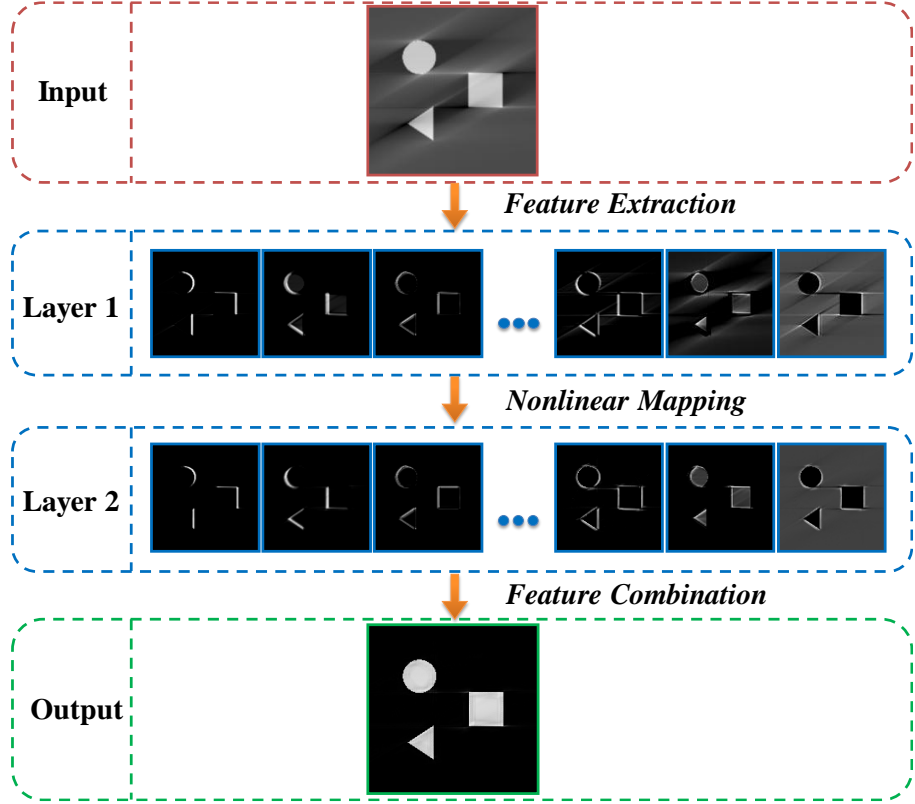

Fig. 3. Basic steps of the proposed network. The input is represented of a series of feature maps in layer 1. Then nonlinear mapping operation maps these feature maps to another ones, and the new feature maps in layer 2 contains fewer characteristics of undesired artifacts. Finally, the output is reconstructed from the new feature maps.

### 2.3.3 Feature combination

The last step is to recombine the feature maps to a whole image. It also could be executed by a convolutional operator and produce the final output by linear combination. The formula of this step can be written as follows:

$$f_3(\mathbf{x}_2) = \mathbf{w}_3 * \mathbf{x}_2 + \mathbf{b}_3, \tag{6}$$

where $\mathbf{w}_3$ and $\mathbf{b}_3$ represent the weight matrix and bias of the convolutional operator, respectively. $\mathbf{w}_3$ is of a size of $C_2 \times n_2 \times n_2 \times 1$, $\mathbf{b}_3$ is a 1-dimensional vector, and $n_2 \times n_2$ is the spatial size of the kernel.

### 2.4 Parameter Selection

Parameters $C_1$ and $C_2$ control the number of extracted features. In general, a larger value of them could decompose and classify more details of the image characteristics, and may result in a better performance. To preserve a balance between performance and computation, a value between 10 and 200 is adequate for most cases. As the network is often designed with a sparser trend in its progressive layers, $C_2 \leq C_1$ is often set.

Parameters $n_1$ and $n_2$ control the spatial supports of the convolutional kernels of step 1 and step 3, respectively. Based on our experience, $n_1 \geq 3$ is preferred for feature extraction, and the value of $n_2$ needn't to be large as the last step is a linear combination operation.

In addition, the setting of layers in nonlinear mapping step is flexible. In our implementation, this step only includes a convolutional layer. In fact, to increase the non-linearity of system function, the strategy uses more convolutional layers with kernel size of $1 \times 1$ may show a better performance.

## 2.5 Experimental studies

To demonstrate and validate our new method for artifact reduction, we performed an experiment on standard clinical data of human body. All the original CT images, with DICOM format, were obtained from the Grassroots DICOM library (http://gdcm.sourceforge.net/wiki/index.php/Sample_DataSet). A geometry representative of a 2D parallel-beam CT scanner setup was used, and the sinogram was simulated by forward projecting the clinical images. The resolution of the CT image was $512 \times 512$ pixels, and each view of simulated sinogram was modeled with 729 bins on a 1D detector.

Three cases with different angular coverage of scanning were considered. Accordingly, three neural networks were trained for different angular cases. The angular ranges of sinogram were set to 170, 150, and 130 degrees for case 1, 2, and 3, respectively.

In the configuration of our network, the parameter $C_1$, $C_2$, $n_1$, and $n_2$ were set to 64, 32, 9, and 9, respectively. To train our network, 3024 slices of CT images were used. The original images without artifacts were used as labels in the training. And the the images reconstructed by FBP algorithm from three different groups of incomplete projections were used as the input images for three corresponding networks. All training work were performed under Caffe framework [38] running on a PC with a GTX 970 GPU.

**2.6 Quantitative image analysis**

To evaluate the image quality quantitatively, the peak signal-to-noise ratio (PSNR) and universal quality index (UQI) [39] are used as measures of the deviations between the predicted images $\mathbf{f}$ and the reference image $\mathbf{f}_{Ref}$.

The PSNR is used to measure the difference between two images. A large value of it suggests small differences to the reference image. The PSNR is defined as follows:

$$PSNR = 10\log_{10}\left(\frac{MAX^2(\mathbf{f}_{Ref})}{\frac{1}{N}\sum_{i=1}^{N}\left|\mathbf{f}_{Ref}(i)-\mathbf{f}(i)\right|^2}\right) \text{ dB}, \quad (7)$$

where $N$ is the total number of pixels in the image.

The mean, variance, and covariance of intensities are defined as follows:

$$\bar{\mathbf{f}}_{Ref} = \frac{1}{N}\sum_{i=1}^{N}\mathbf{f}_{Ref}(i), \quad \sigma_{Ref}^2 = \frac{1}{N-1}\sum_{i=1}^{N}\left(\mathbf{f}_{Ref}(i)-\bar{\mathbf{f}}_{Ref}\right)^2, \quad (8)$$

$$\bar{\mathbf{f}} = \frac{1}{N}\sum_{i=1}^{N}\mathbf{f}(i), \quad \sigma^2 = \frac{1}{N-1}\sum_{i=1}^{N}\left(\mathbf{f}(i)-\bar{\mathbf{f}}\right)^2, \quad (9)$$

$$Cov(\mathbf{f}_{Ref},\mathbf{f}) = \frac{1}{N-1}\sum_{i=1}^{N}\left(\mathbf{f}_{Ref}(i)-\bar{\mathbf{f}}_{Ref}\right)\left(\mathbf{f}(i)-\bar{\mathbf{f}}\right), \quad (10)$$

Then, the UQI can be calculated as follows:

$$UQI = \frac{2Cov(\mathbf{f}_{Ref},\mathbf{f})}{\sigma_{Ref}^2+\sigma^2}\frac{2\bar{\mathbf{f}}_{Ref}\bar{\mathbf{f}}}{\bar{\mathbf{f}}_{Ref}^2+\bar{\mathbf{f}}^2}. \quad (11)$$

A UQI value closer to one indicates a higher degree of similarity between the predicted image and the reference image.

## 3. Results

To validate and evaluate the performance of the proposed strategy, three slices of clinical data of human body at different positions (head, abdomen, and chest) were used in the experiments. As mentioned before, we had trained three networks for three different limited angle configurations with an angular range of 170, 150, and 130 degrees, respectively. In the experiment of each slice, three different configurations same as the settings of networks were considered. Initial images were generated by FBP method, and used as input images for our networks.

Images corrected and predicated by proposed method of a head slice are displayed in Fig. 4. To reveal texture details, the zoomed ROI images near the left ear are also shown in Fig. 4. The input FBP images suffer from streak artifacts and distortions in different degrees which is consistent to the different conditions of missing angular data. The predicted images show a visible suppression of the streak artifacts and better preservation of details. In addition to quantitative inspection of the results, the PSNR and UQI of images are calculated, and the calculation results are listed in Table 1. The quantitative studies indicate the superiority of the gains from the proposed method in terms of the measurement of image quality. And the results suggest that compared with the input images, the predicted images can achieve an obvious better accuracy that match the references.

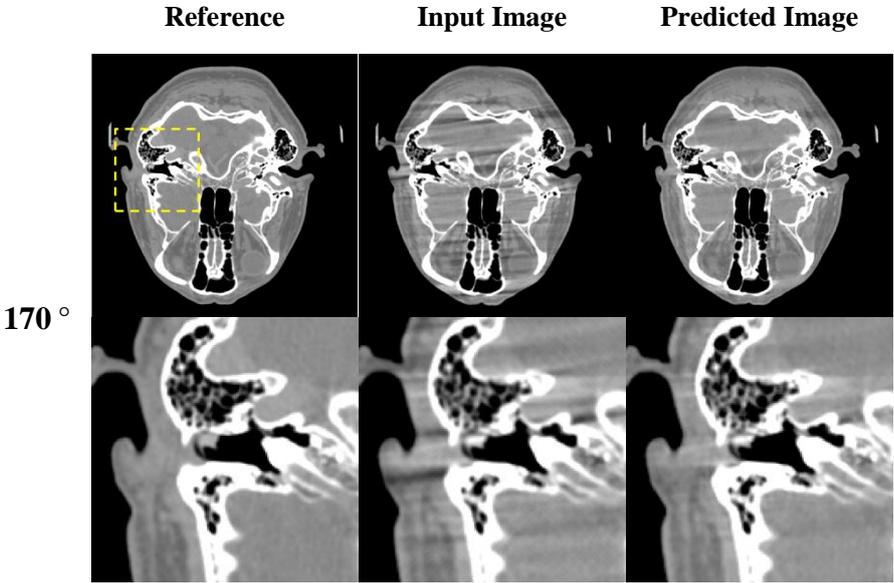

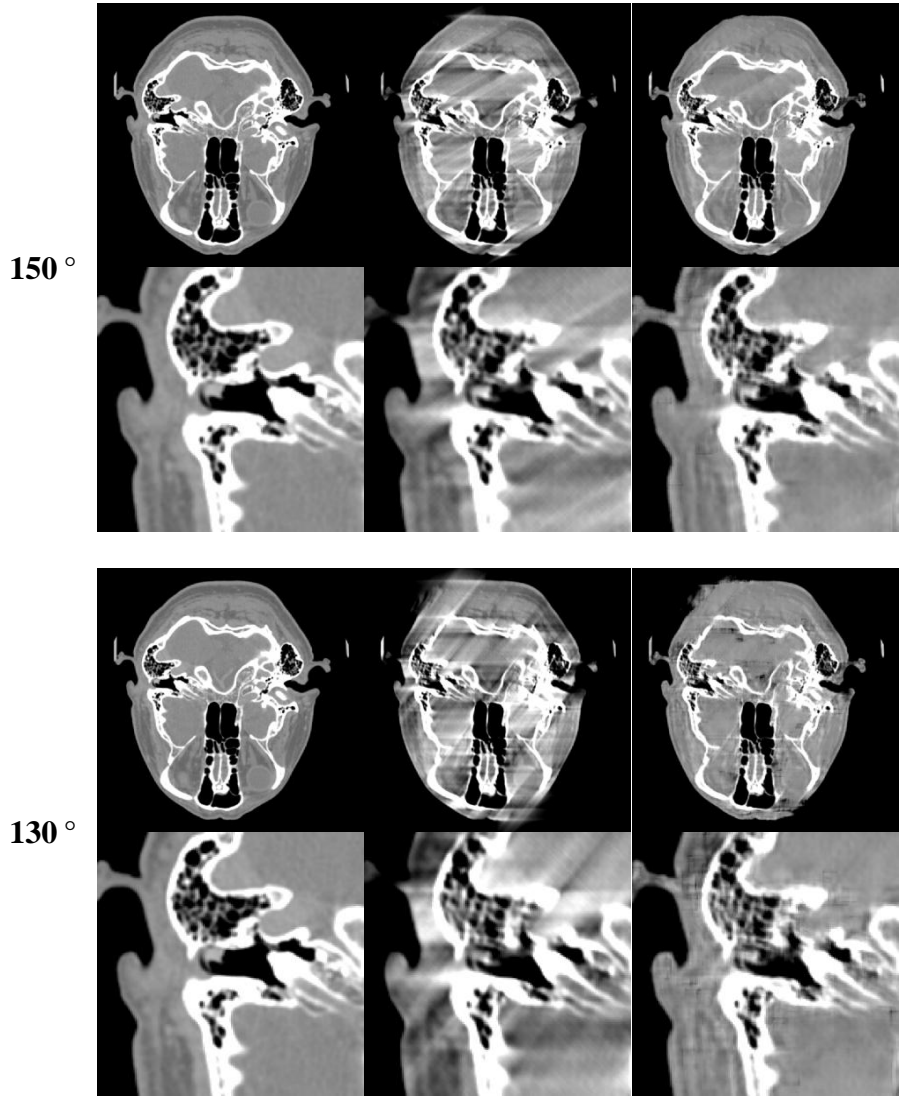

Fig. 4. Image predications in the 170 °(rows 1-2), 150 °(rows 3-4), and 130 °(rows 5-6) scanning cases of a slice of head CT data. From left to right in each row, images of the references, input FBP images, and predicted images are presented. The display window is [-500HU, 500HU].

Table 1. Evaluations of the results predicted by different networks in the head data study.

|  |  | Input Image | Predicted Image |
|---|---|---|---|
| 170 ° | PSNR | 30.95 dB | 35.36 dB |
|  | UQI | 0.99118 | 0.99657 |
| 150 ° | PSNR | 24.13 dB | 30.66 dB |
|  | UQI | 0.95704 | 0.98851 |
| 130 ° | PSNR | 20.91 dB | 26.30 dB |
|  | UQI | 0.89902 | 0.97412 |

Figure 5 compares the artifacts correction performances of 170 °and 150 °scanning in an abdominal data study. The uncorrected FBP images suffer from streak artifacts in all cases.

The proposed ConvNet-based method could remove streak artifacts successfully, though some small distortions still exist in the direction of missing data in the case of 150°scanning. The PSNR and UQI of the abdominal CT images are given in Table 2. The quantitative results from the proposed method showed noticeable gains in terms of the two measures.

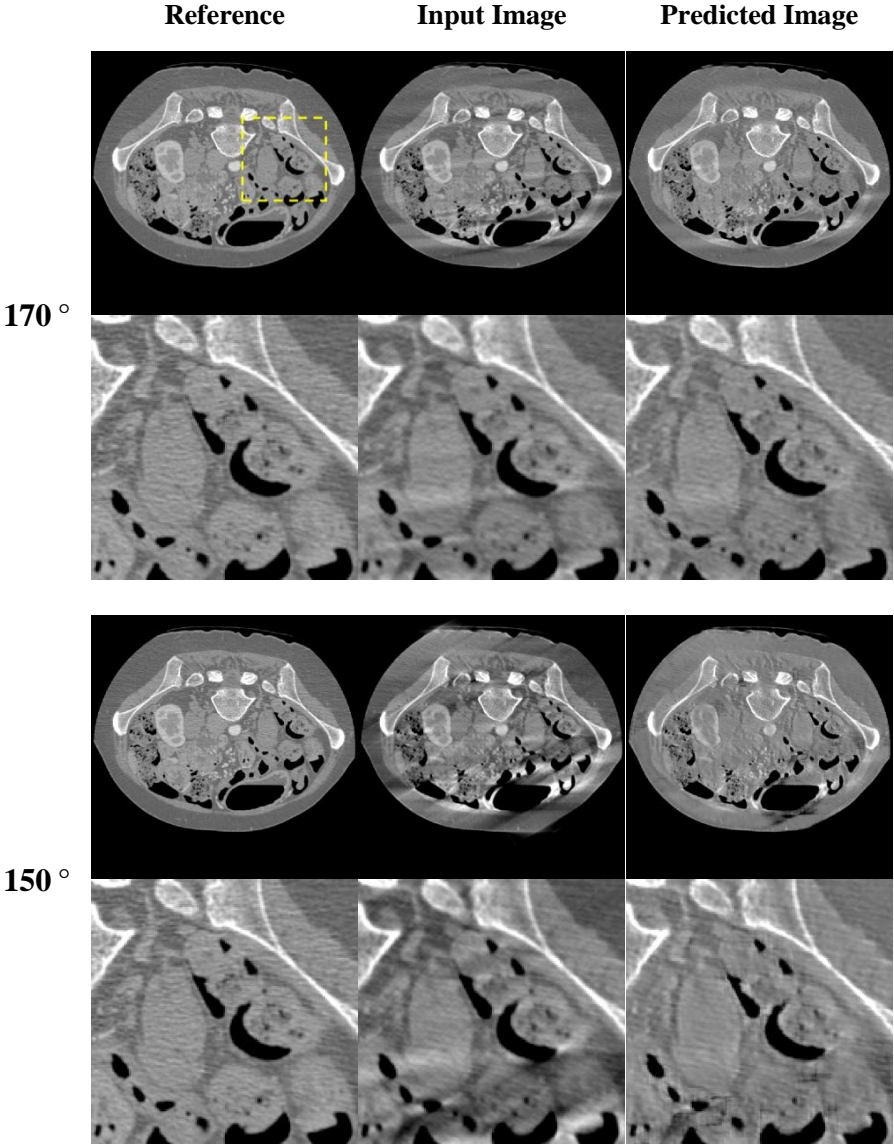

Fig. 5. Image predications in the 170°(rows 1-2) and 150°(rows 3-4) scanning cases of a slice of abdominal CT data. From left to right in each row, images of the references, input FBP images, and predicted images are presented. The display window is [-500HU, 500HU].

Table 2. Evaluations of the results predicted by different networks in the abdominal data study.

|  |  | Input Image | Predicted Image |
|---|---|---|---|
| 170° | PSNR | 29.70 dB | 35.04 dB |
|  | UQI | 0.98736 | 0.99628 |
| 150° | PSNR | 23.47 dB | 30.50 dB |
|  | UQI | 0.94145 | 0.98911 |
| 130° | PSNR | 20.63 dB | 25.81 dB |
|  | UQI | 0.87029 | 0.96964 |

The corrected images and zoomed-in images corresponding to the selected ROIs of a chest data slice from 170° and 150° scans are presented in Fig. 6. As one can see, streak artifacts exist in the FBP results, and the proposed ConvNet-based method yield noticeable performance in terms of streak artifact suppression and deformities correction. To further display the gains of the proposed method, the PSNR and UQI of each image were also calculated, and the results are listed in Table 3. The results show that our method exhibit a prospective performance in terms of accuracy and resolution properties, which agrees with the findings in Table1 and Table 2.

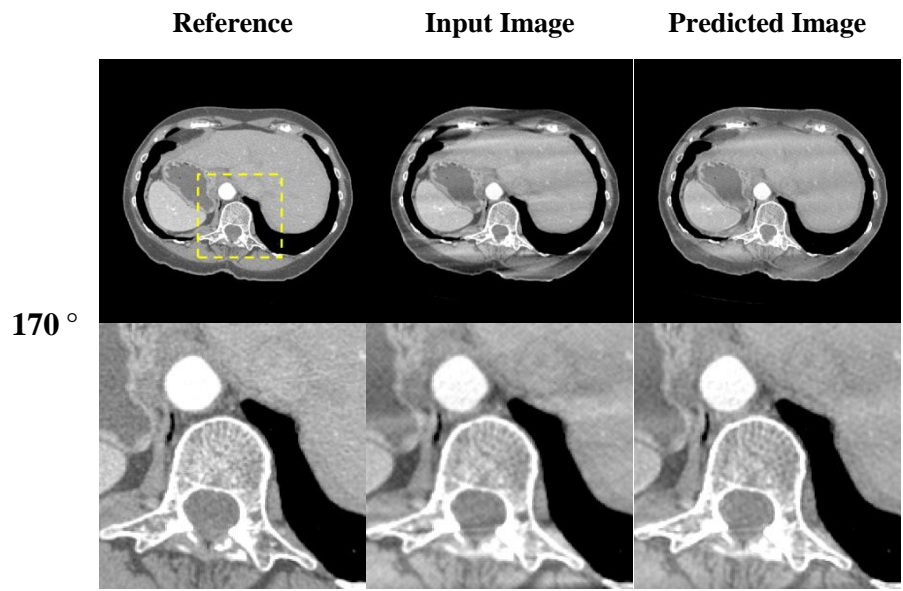

170°

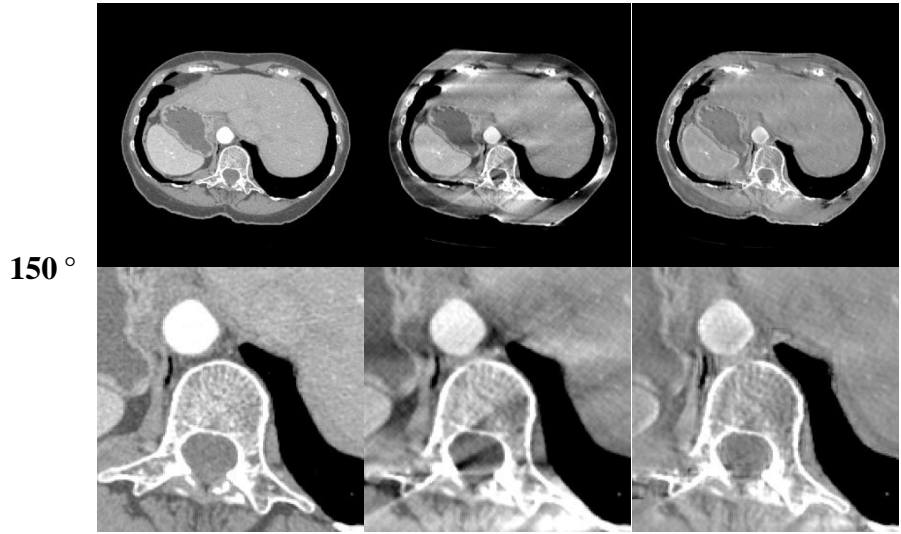

Fig. 6. Image predications in the 170°(rows 1-2) and 150°(rows 3-4) scanning cases of a slice of chest CT data. From left to right in each row, images of the references, input FBP images, and predicted images are presented. The display window is [-300HU, 300HU].

Table 3. Evaluations of the results predicted by different networks in the chest data study.

|  |  | Input Image | Predicted Image |
|---|---|---|---|
| 170° | PSNR | 29.79 dB | 35.92 dB |
|  | UQI | 0.98859 | 0.99688 |
| 150° | PSNR | 22.98 dB | 31.24 dB |
|  | UQI | 0.94513 | 0.99147 |
| 130° | PSNR | 19.54 dB | 26.74 dB |
|  | UQI | 0.87560 | 0.97825 |

The average running time of the whole convolutional process on an image with a size of 512×512 in the experiments is about 8.3 seconds under MATLAB 2012a running on a PC with an Intel I7-3770 3.40 GHz CPU. And the time cost of training procedure of the network is about one week on a GTX 970 GPU. Though the training is very time consuming, the trained result could be used for a stationary appliance with a fast access. And the computation time of image predication procedure is much less than that of iterative reconstruction methods. In particular, this procedure also could be accelerated evidently by graphics processing units.

## 4. Discussion and Conclusion

Limited angle problem is an open problem in x-ray CT field. Due to the deficiency in continuous angular data, the reconstructed image of standard FBP method is deteriorated by serious artifacts. In this study we present a novel deep learning approach for reducing these limited angle artifacts in FBP results. The proposed method considers the extraction and suppression on implicit features of the specific artifacts from original FBP images, and it shows an excellent performance on artifact suppression and feature preservation.

The new strategy can improve the FBP image quality remarkably and would be helpful for clinical diagnosis as it could be integrate into practical applications immediately with only little increase of computation. In addition, current mainstream mean for image reconstruction from incomplete sinogram is iterative reconstruction method. We show that a deep learning-based "FBP + Artifacts Reduction" method could also provide a similar satisfied result. This experience should broaden our understanding of the image reconstruction issue and provide a new practical solution to incomplete data reconstruction.

The main problem of our method is that it needs a large dataset for training and the training procedure often requires huge amounts of computing resources. With the development of high-performance devices, an advisable method for reducing computational cost is to implement the training with the acceleration of the distributed computing systems. Besides, the proposed method couldn't be used flexibly and suffers limitations in application. It should be applied in a fixed scanning configuration which is consistent to the condition of the trained dataset.

When a large angular coverage of projection data is not acquired, the FBP reconstructions will miss much information and the details will be blurred seriously. In this case, our method will not work well. A network which not only considers the FBP reconstructions as input but also builds a relationship between sinograms and predicted images may be a promising solution to produce high-quality images for limited angular reconstruction. Addressing this question is one of our future research focuses.

The structure of our network is flexible. The performance of it could be further improved via increasing the scale of datasets or exploring more hidden layers of the network, and the structure of it may also be applied to some other applications, e.g. sparse-view CT image reconstruction, noise suppression in low-dose CT.

In conclusion, this paper proposes a deep ConvNet-based method to reduce the artifacts in images reconstructed by FBP method from limited angular projections. The features involving complex artifacts will be extracted and restrained by nonlinear mapping approach, thus an image with fewer artifacts will be predicted in the output of our network. The experimental results demonstrated the performance of the new method in artifact suppression and image quality improvement. The findings in this paper enabled a more effective artifact reduction approach and suggested a potential use for clinical diagnosis.

## Acknowledgements

This work was partially supported by the grants from NSFC 61372172, 61571256 and 81427803.